\documentclass[conference]{IEEEtran}
\IEEEoverridecommandlockouts
\usepackage{cite}
\usepackage{amsmath,amssymb,amsfonts}
\usepackage{algpseudocode}
\usepackage{graphicx}
\usepackage[utf8]{inputenc}
\usepackage{siunitx}
\usepackage{booktabs, makecell}
\ifCLASSOPTIONcompsoc
    \usepackage[caption=false, font=normalsize, labelfont=sf, textfont=sf]{subfig}
\else
\usepackage[caption=false, font=footnotesize]{subfig}
\fi
\usepackage{nomencl}
\usepackage{multirow,tabularx}
\usepackage{etoolbox}
\usepackage{textcomp}
\usepackage{xcolor}
\usepackage{verbatim}
\usepackage{mathtools}
\usepackage{mdwtab}
\usepackage{array}

\makeatletter
\newcommand*{\rom}[1]{\expandafter\@slowromancap\romannumeral #1@}
\makeatother

\begin{document}

%\title{Pre-trainined Vision Transformer Model \\for Load Image Understanding}

%\title{Vision Transformers for Understanding Load Images: A Paradigm Shift in Load Profile Analysis}

%\title{Pre-trained Vision Transformer Model for Understanding Load Images}
\title{A Novel Vision Transformer based Load Profile Analysis using Load Images as Inputs}
%\title{Innovative Load Profile Analysis through Vision Transformer with Load Image Inputs}

\author{\IEEEauthorblockN{Hyeonjin Kim, Yi Hu, Kai Ye, Ning Lu}
\IEEEauthorblockA{\textit{North Carolina State University} \\
\textit{Raleigh, NC 27606, USA}\\
\{hkim66, yhu28, kye3, nlu2\}@ncsu.edu}
}

\maketitle

\begin{abstract}
This paper introduces ViT4LPA, an innovative Vision Transformer (ViT) based approach for Load Profile Analysis (LPA). We transform time-series load profiles into load images. This allows us to leverage the ViT architecture, originally designed for image processing, as a pre-trained image encoder to uncover latent patterns within load data. ViT is pre-trained using an extensive load image dataset, comprising 1M load images derived from smart meter data collected over a two-year period from 2,000 residential users. The training methodology is self-supervised, masked image modeling, wherein masked load images are restored to reveal hidden relationships among image patches. The pre-trained ViT encoder is then applied to various downstream tasks, including the identification of electric vehicle (EV) charging loads and behind-the-meter solar photovoltaic (PV) systems and load disaggregation. Simulation results illustrate ViT4LPA's superior performance compared to existing neural network models in downstream tasks. Additionally, we conduct an in-depth analysis of the attention weights within the ViT4LPA model to gain insights into its information flow mechanisms.

%We leverage the ViT architecture as a pre-trained image encoder, utilizing its specialized transformer design tailored for image processing tasks. 
%In the pre-training stage, the ViT model processes corrupted load images where a few of the image patches are masked. The ViT model is trained given an image-recovering task where the model outputs a reconstructed image using only partial information of the original load image. Through this pre-training, the ViT model can learn the in-depth latent representation of load image by inferring the relationship between image patches. This pre-trained encoder can be utilized for various customer-level downstream tasks and we present the PV and electric vehicle (EV) detection as one example. Simulation results show that the proposed strategy outperforms existing neural network models on PV and EV detection tasks. Furthermore, we enhance the understanding of the ViT model by analyzing attention weights to speculate about its detection mechanism. 

\end{abstract}


\section{Introduction}

Pre-trained neural network models have been widely used in Natural Language Processing (NLP)~\cite{devlin2018bert} and Computer Vision (CV)~\cite{he2022masked} tasks. Compared to training a model from scratch, using pre-trained models reduces reliance on labeled data and saves considerable time and computational resources when performing downstream tasks~\cite{radford2018improving}. In the field of natural language processing (NLP), the adoption of large pre-trained language models, such as BERT and GPT, has brought about a transformative shift in the domain of language comprehension and generation. These models undergo extensive training on extensive text corpora, endowing them with the capacity to apprehend intricate linguistic structures and contextual nuances. Consequently, NLP has undergone remarkable progress in the past couple of years, with pre-trained models emerging as pivotal cornerstones for a diverse array of downstream applications, including but not limited to sentiment analysis, machine translation, text summarization, and question answering. 

The transformer model exhibits exceptional scalability, surpassing traditional machine learning based models such as Recurrent Neural Networks (RNNs) and convolutional neural networks (CNNs). This led to the creation of the Vision Transformer (ViT) ~\cite{dosovitskiy2020image}, specifically designed for processing computer vision (CV) tasks. In contrast to conventional CNNs, ViT uses a transformer architecture by dividing images into fixed-size patches, linearly projecting them, and employing self-attention mechanisms to capture long-range dependencies among patches. ViT soon excels in conducting a wide variety of CV tasks, including image classification, object detection, and segmentation. The success of ViT in the realm of CV underscores the transformative potential of applying transformer architectures across different domains.

In the field of power system analysis, load profile analysis (LPA) is becoming increasingly vital for performing tasks like load disaggregation, model parameterization, customer segmentation, load flexibility analysis, and identifying behind-the-meter resources. However, a distinctive challenge in this field is the scarcity of publicly accessible, non-sensitive power system datasets. The sensitivity and proprietary nature of power grid operations make it not only financially burdensome but often impractical for researchers and developers to amass the substantial volume of training data required for training resilient machine learning models.

\begin{figure*}
	\centering
	\includegraphics[width=6.9in]{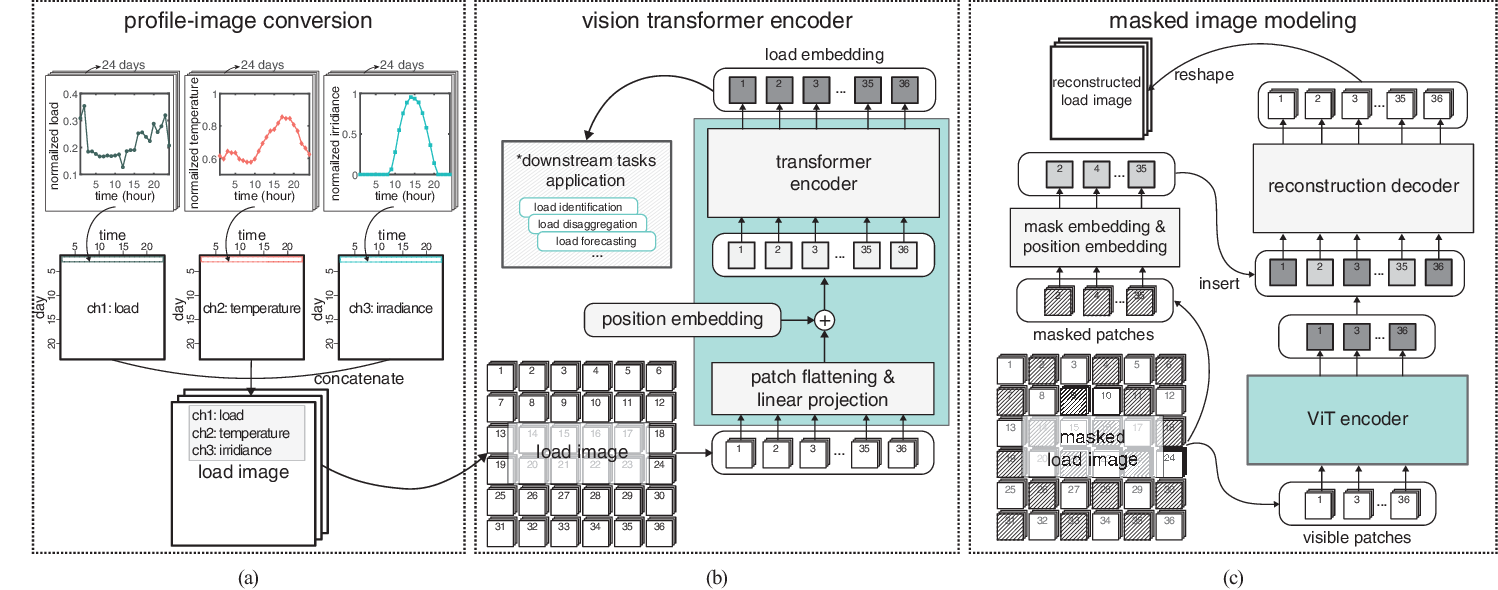}
\caption{An illustration of the ViT4LPA architecture. (a) Profile-to-image conversion, (b) ViT4LPA workflow, and (c) Pre-training process.}
\label{Vit_Scheme}
\vspace{-0.2in}
\end{figure*}

Hence, leveraging extensive data for pre-training a model applicable to subsequent LPA tasks, akin to the use of BERT in NLP and ViT in CV, can effectively address the challenge of restricted data accessibility.
%Therefore, in this paper, we propose ViT4LPA, an innovative Vi-sion Transformer (ViT) based approach for Load Profile Analysis (LPA). By converting time-series load profiles into load images, ViT4LPA is pre-trained using an extensive load image dataset derived from smart meter data.The training methodology is self-supervised, masked image modeling, wherein masked load images are restored to reveal hidden relationships among image patches. The pre-trained ViT encoder is then applied to various down-stream tasks, including the identification of electric vehicle (EV) charging loads and behind-the-meter solar photovoltaic (PV) systems and load disaggregation. }
To date, the development of a versatile pre-trained model suitable for various downstream tasks in the field of LPA has not received adequate attention from researchers in the power system domain. Recent research endeavors have prominently revolved around transfer learning, a technique wherein neural networks, initially trained for specific supervised learning tasks, are repurposed to perform related tasks by leveraging the knowledge they have accrued. In~\cite{d2019transfer}, cross-domain transfer learning was introduced, aiming to apply latent features learned from one appliance to another, particularly for non-intrusive load monitoring tasks. In~\cite{ryu2019convolutional}, the authors explored LPA through representation learning using convolutional autoencoders. In our recent work~\cite{hu2023bert}, we introduced a BERT-based load profile inpainting approach, serving as a foundational model to streamline the restoration of missing data tasks. Nonetheless, it is currently primarily focused on addressing a specific LPA task.

Thus, in this paper, we present a pre-trained model framework tailored for LPA. Our main contribution is the introduction of ViT4LPA, a novel approach based on Vision Transformers (ViT) specifically designed for LPA tasks. This approach marks a significant shift in the LPA domain by not only reducing the dependence on labeled datasets but also enabling its seamless application across a diverse spectrum of downstream tasks.

\section{Methodology} \label{Methodology}
This section introduces ViT4LPA, an innovative Vision Transformer (ViT) based approach for LPA.

\subsection{Profile-Image Conversion} \label{Vit}
One of our key contributions is to base the LPA study on load images derived from these time-series load profiles. Traditional LPA studies typically use time-series load profiles as inputs to capture temporal cyclic characteristics, including daily, weekly, or monthly variation patterns. 
However, in a more recent development, a few authors have proposed the conversion of load profiles into color-coded load images for autoencoder-based clustering, as seen in \cite{ryu2019convolutional}, and for BERT-based missing data recovery, as demonstrated in \cite{hu2023bert}. Motivated by the accomplishments in those endeavors, we use load images, as opposed to load profiles, to pre-train a ViT-based foundational model that can be fine-tuned with a small amount of data when conducting downstream LPA tasks.
%we use load images, rather than load profiles, for pretraining a ViT-based foundational model for subsequent LPA tasks.

As illustrated in Fig.~\ref{Vit_Scheme}(a), we generate load image comprising three distinct channels. Channels 1 to 3 encompass data from smart meter loads, temperature readings, and irradiance profiles, respectively. This approach allows the encoder to capture not only hidden patterns within the load data but also the correlation with temperature and irradiance variations. $x-$axis of the image corresponds to the number of data ($N_T$) within one day and $y-$axis represents the number of days ($N_D$). To convert a data point into a color patch, we initially perform a linear projection of the data point onto the [0,1] range. For instance, consider the case where the minimum load consumption ($p^-$) is $-$4kW and the maximum load consumption ($p^+$) is 24kW. In this scenario, the load consumption at hour $t$, ${p_t}$, can be normalized as follows: $(p_t - p^-)/(p^+ - p^-)$.

Load images excel at condensing vast amounts of multi-modal data, encompassing variables like load, temperature, and solar irradiance, facilitating the streamlined consolidation of information. This proves especially advantageous when compressing monthly or yearly data into load images, aligning profiles based on their cyclical characteristics. Through the transformation of load profiles into load images, we can seamlessly apply the ViT model \cite{dosovitskiy2020image}, initially designed for computer vision tasks, to effectively handle LPA tasks.

%When an intra-day load profile is used as an input the conventional model selection has been a sequence transduction model including models with RNNs and 1D convolution layers. This comes from similar highly semantic and information-dense properties between load profile and language. However, when the load profile including multiple days is analyzed its repetitive pattern reveals spatial redundancy, which is a property of images. In~\cite{ryu2019convolutional}, Ryu \textit{et al.} constructs the yearly load profile as a load image, and its redundant information is compressed through the feature extraction model. Based on a similar strategy, this paper reformulates the load profile into a load image, and by leveraging the ViT model, we aim to build the pre-trained model that extracts customer-specific representation called \textit{load embedding}.

\subsection{Vision Transformer Encoder Inputs}
In Fig.~\ref{Vit_Scheme}(b), we present the workflow of ViT4LPA. To directly apply the ViT model architecture, we initially partition a load image comprising $N_D\times N_T$ pixels into $\frac{{N_D}\times{N_T}}{{N_P}^2}$ image patches, ensuring each patch contains $N_P\times N_P$ pixels. Subsequently, these image patches are flattened into a sequence of patches, ready to be fed as inputs to the ViT model. 

Note that the division of load patches is contingent on both the data resolution ($N_T$) and data duration ($N_D$). 
In this paper, due to the page limit, we will focus on introducing the ViT4LPA architecture and workflow.  
The detailed discussion regarding the selection of $N_P$ in relation to $N_T$ and $N_D$ will be presented in our follow-up journal paper. Thus, we fix ${N_T}=24$ hours, ${N_D}=24$ days, and $N_P=4$. In this setup, a load image will contain 24 rows (representing the number of days) and 24 columns (representing the number of data points in a day). After partitioning into $6\times 6$ image patches, each patch contains $4\times 4$ color patches. We then flatten the 36 image patches and send them directly to the ViT encoder (the blue block in Fig.~\ref{Vit_Scheme}(b)) for processing. The descriptions of the ViT encoder model can be found in~\cite{dosovitskiy2020image}.

The load embeddings produced by the pre-trained ViT encoder can be effectively used for various downstream LPA tasks. In this paper, we use the behind-the-meter load identification task and load disaggregation task as examples to showcase the efficacy of ViT4LPA. ViT4LPA will be employed to identify the following three types of distributed energy resources: electric vehicle (EV) charging load, Photovoltaic (PV) generation, and Heating, Ventilation, and Air-Conditioning (HVAC) load.

\subsection{Pre-training ViT4LPA using Masked Image Modeling Tasks}
In Fig.\ref{Vit_Scheme}(c), we describe the architecture of the masked autoencoder network used to pre-train the ViT4LPA encoder, following the masked autoencoder scheme introduced in\cite{he2022masked}. The training task involves restoring the masked image patches using a reconstruction decoder. The input to this decoder is the load image embedding produced by the ViT encoder using unmasked image patches as inputs. In Fig. 1(c), since the ViT-generated embeddings are for the unmasked image patches, it is necessary to incorporate the mask embeddings to ensure that the input of the decoder matches the dimensionality of the original load image.

As depicted in Fig.~\ref{Vit_Scheme}(c), 18 image patches are subjected to masking, leading to the ViT encoder receiving only the visible 18 image patches as input. Consequently, the output of the ViT encoder consists of only 18 embeddings, requiring the insertion of 18 mask embeddings to achieve the necessary 36 embeddings for the reconstruction decoder. The 36 restored image patches by the decoder will be rearranged back to a load image and compared with the original load image to calculate losses.
%Only the visible part of the image $x_p^*\in {\mathbb{R}^{{N^*} \times 3{P^2}}}$ is used as an input for ViT encoder where $N^*$ is a number of visible patches. 
%By reconstructing the missing patches, we train a masked autoencoder is to reconstruct the missing patches within the pixel space. 

Please note that various masking strategies, including grid-masking, random-masking, span masking, and more, exist. However, due to page limitations, this paper exclusively introduces the grid-masking approach. This choice aligns with the primary focus of the downstream LPA tasks discussed in this paper, i.e., the tasks of behind-the-meter load identification and load disaggregation. In these tasks, grid masking has been found to demonstrate superior training efficiency when compared to random masking. Consequently, for this paper's scope, we will exclusively showcase results obtained through grid masking. A more in-depth examination of the impacts of different masking strategies (e.g., random masking or variable grid masking) on downstream task performances will be extensively addressed in our follow-up journal paper.

\section{Simulation Results}
In this section, we showcase the efficacy of the proposed pre-trained ViT4LPA model in three downstream load identification tasks, using smart meter data as inputs.

\subsection{Simulation Setup}

To construct the training dataset for the ViT encoder, we generate 4,000 sets of 1-hour resolution yearly load profiles. The original datasets are collected by the Pecan Street project~\cite{pecan}. The load profiles are metered from 150 households in Austin, Texas, including sub-metered PV, EV, and HVAC load consumption. The dataset spans 2 years, with a 1-minute data resolution. To streamline the load identification task, we downsample the 1-minute data to 1 hour. 
The data augmentation strategy introduced in~\cite{ye2023modified} is used to generate additional training examples for bolstering the model's resilience. 
The dataset is partitioned as follows: the training set comprises load profiles from 100 users, while the testing dataset includes the remaining 50 customers. This ensures a comprehensive evaluation of the proposed models.

%All load profiles are normalized by using minimum and maximum load consumption over all of the load profiles. Minimum ($\underset{\raise0.3em\hbox{$\smash{\scriptscriptstyle-}$}}{p}$) and maximum load consumption ($\bar p$) are selected to be 4kW and 24kW, respectively. Then, the original load consumption value at each hour (${p_t}$) is normalized into the 0 to 1 range calculated by $(p_t+\underset{\raise0.3em\hbox{$\smash{\scriptscriptstyle-}$}}{p})/\bar p$.

\subsection{Performance on the Masked Image Modeling Tasks} \label{sectionIIIB}
\begin{figure}[h]
	\centering

 \subfloat[]{\includegraphics[width=3.2in]{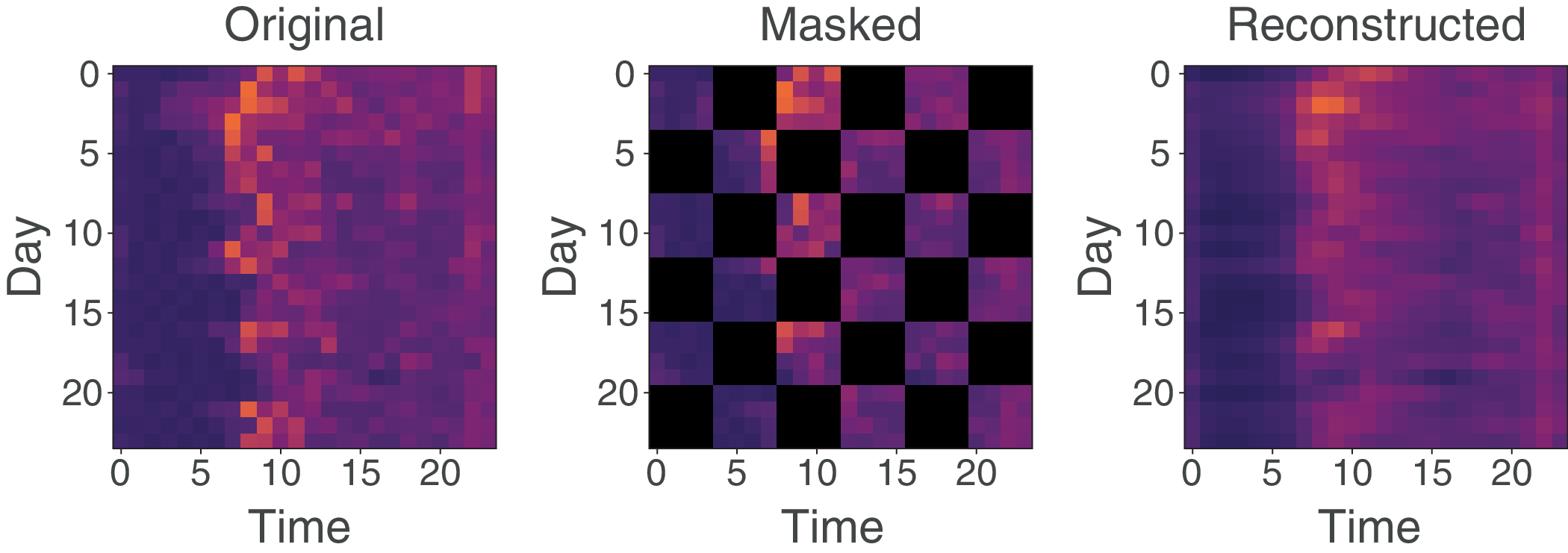}\label{pos}}

\subfloat[]{\includegraphics[width=3in]{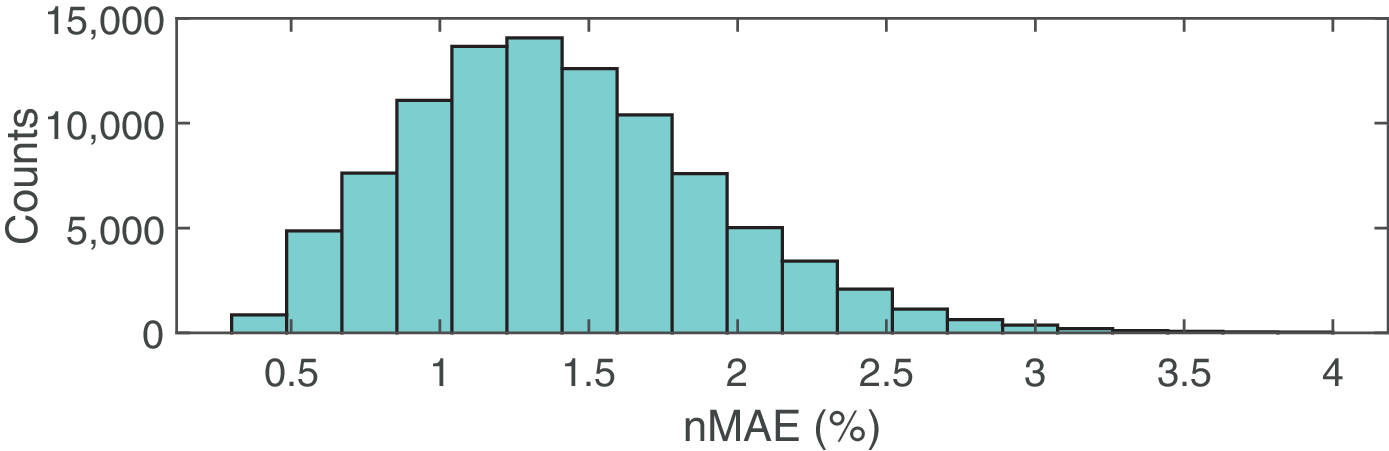}\label{reconst}}
\caption{ ViT4LPA encoder performance evaluation. (a) An illustration of the original, masked, and reconstructed load image, and (b) Reconstruction error distribution. }
\label{Recon}
\vspace{-0.1in}
\end{figure}

%In this section, we present the simulation results obtained from the masked autoencoder. 
The hyperparameters used for pre-training the ViT4LPA encoder model (see the network architecture depicted in Fig.~\ref{Vit_Scheme}(c)) are listed in Table~\ref{MAE_Detail}. As shown in Fig.~\ref{Recon}(a), the reconstruction results demonstrate that the model can reconstruct the load image from partial information (i.e., unmasked image patches) with satisfactory performance. Interestingly, the masked autoencoder demonstrates increased effectiveness in recovering missing patches during the morning to noon period (8:00-12:00), which aligns with the typical peak consumption hours for the specific customer. In Fig.~\ref{Recon}(b), $n$MAE distribution of the masked image reconstruction is represented. Error is concentrated in the 1-2\% range, where the mean and standard deviation of the error are 1.40\% and 0.515\%, respectively.

\begin{table}[h]
	\begin{center}
		\caption{Parameters Used in the ViT4LPA Pre-training Process}
		\label{MAE_Detail}
		\begin{tabular}{ccccc}
	
			\toprule % <-- Toprule here
Network & Layers & Heads  & Proj. Dim. ($D$) & Parameters\\

\midrule

\textbf{Encoder} & 3 & 4 & 128 & 1.0M \\
\textbf{Decoder} & 2 & 2 & 32 & 2.0M\\
\midrule
 \multirow{ 2}{*}{Hyperparam.}& Batch size & Dropout & Optimizer & Epochs\\
\cmidrule{2-5}	
& 64 & 0.1 & Adam & 50\\
\midrule
\end{tabular}
	\end{center}
\vspace{-0.2in}
\end{table}

\vspace{-0.1in}

%As depicted in Fig.~\ref{Recon}, the reconstruction outcome of the masked autoencoder for the masked image modeling task is showcased. The results underscore the model's proficiency in deciphering load consumption patterns based on partial information. Notably, the masked autoencoder effectively recovered missing patches, particularly during the morning to noon period (8:00-12:00), which aligns with the typical peak consumption hours for the specific customer under consideration. 

%\subsection{Interpretation of ViT Encoder Network}

\begin{figure}[h]
	\centering
	\includegraphics[width=2.8in]{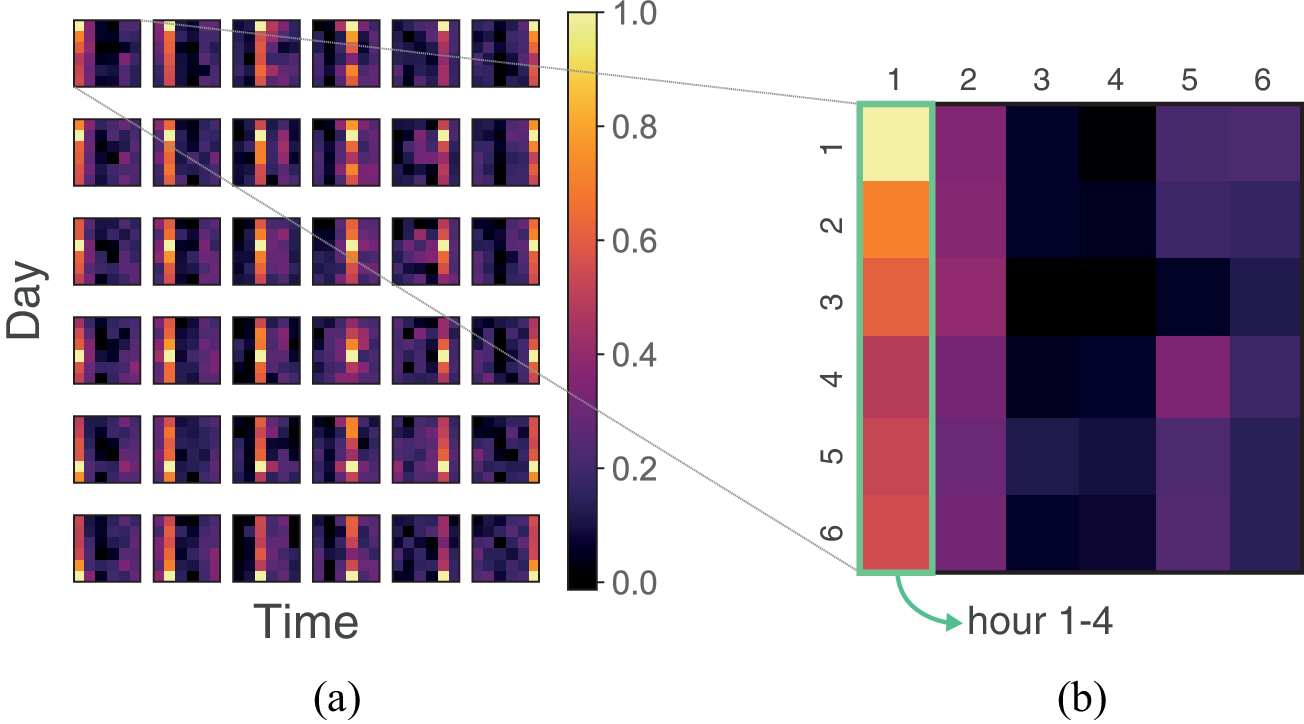}
\caption{Analysis on position embeddings of the ViT4LPA encoder. (a) Heatmaps of the 36 similarity matrices and (b) Similarity matrix of image patch 1.}
\label{Fig3}

\end{figure}

Subsequently, the cosine similarity matrix is employed to assess the similarity of position embeddings among patches. In Fig.\ref{Fig3}(a), we depict heatmaps showcasing 36 similarity matrices. Each matrix corresponds to the cosine similarity of position embeddings between an image patch at a specific position in the load image, in relation to the 36 image patches, which include itself. To illustrate, in Fig.\ref{Fig3}(b), the top-left similarity matrix represents the cosine similarity of position embeddings between the initial image patch (consisting of 48 data points from day 1-4 and hour 1-4) and the remaining 36 image patches.

Notably, when considering photo image patches, we observe significant similarities between a patch and its neighboring patches, regardless of their alignment in columns, diagonally, or rows, as discussed in~\cite{dosovitskiy2020image}. In contrast, for load image patches, high similarity is primarily found among patches aligned in columns, indicating strong correlations among profiles at the same time of day across different days. For instance, data from hours 1-4 exhibits a strong correlation with data from hours 1-4 on all other days (i.e., in the column-wise direction). However, this correlation may be less prominent when comparing data from adjacent hours, such as between hours 1-4 and hours 5-8 (i.e., along the row-wise direction).

\begin{figure}[h] 
\centering
%\subfloat[position embedding]{\includegraphics[width=1.7in]{Figure/pos_emb.eps}\label{pos}}
%\hspace{0.1em}
\subfloat[]{\includegraphics[width=1.15in]{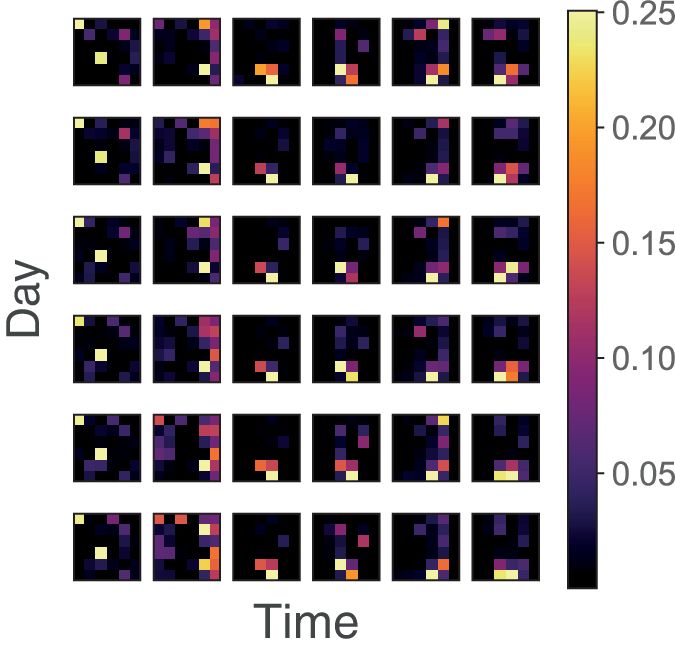}}
%\hspace{0.1em}
\subfloat[]{\includegraphics[width=1.15in]{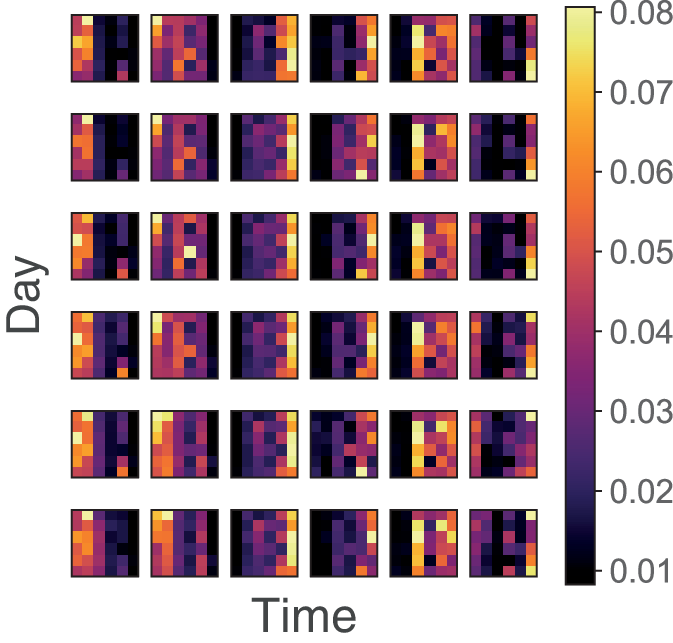}}
%\hspace{0.1em}
\subfloat[]{\includegraphics[width=1.15in]{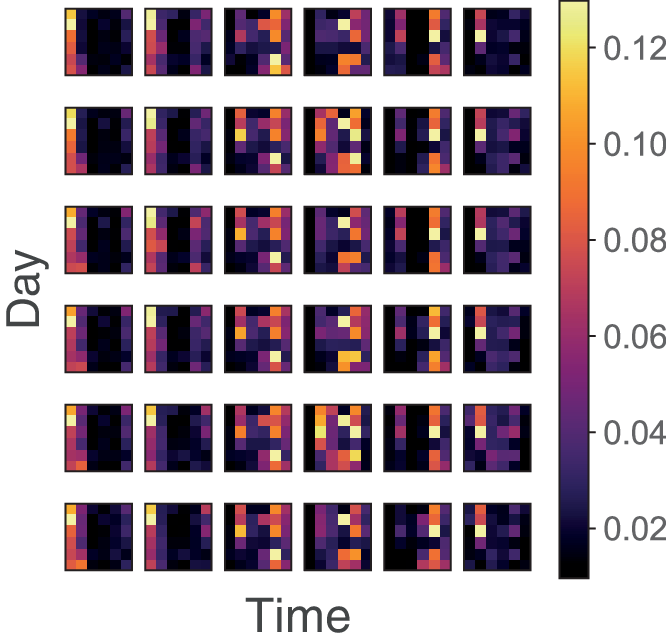}}\\
\caption{Heatmaps of the mean of the self-attention layers. (a) Layer 1, (b) Layer 2, and (c) Layer 3.}
\label{Fig4}
\vspace{-0.2in}
\end{figure}

In Figs.~\ref{Fig4}(a) to (c), the mean self-attention matrices, averaged across multiple attention heads within each layer, are depicted. Similar to Fig.~\ref{Fig3}(a), the 1-by-36 attention weights are reshaped into a 6-by-6 matrix to incorporate the positional information of patches. Analyzing the mean attention matrices provides valuable insights into the local and global information flow within each layer and the receptive field of the model. An important insight obtained from the study is that in the first layer, the attention weights are sharply focused on a small set of patches, indicating a localized focus on specific details. As the network deepens, particularly in layers 2 and 3, attention is dispersed across a larger number of patches compared to the initial layer. This shift implies the model's transition from capturing localized features to embracing a more holistic perspective, encompassing the entire load image to extract comprehensive global information.

\subsection{Downstream Task 1: Load Identification}
To demonstrate the effectiveness of using pre-trained model on downstream tasks, we first compare the performance of ViT4LPA on two load identification tasks: behind-the-meter PV and EV identification. 
The identification models have the same network structure for both the scenarios with and without ViT4LPA. 
Note that the ViT4LPA encoder model can be further fine-tuned using a small amount of labeled training data.

\begin{figure}[h]
\centering

\subfloat[]{\includegraphics[width=1.5in]{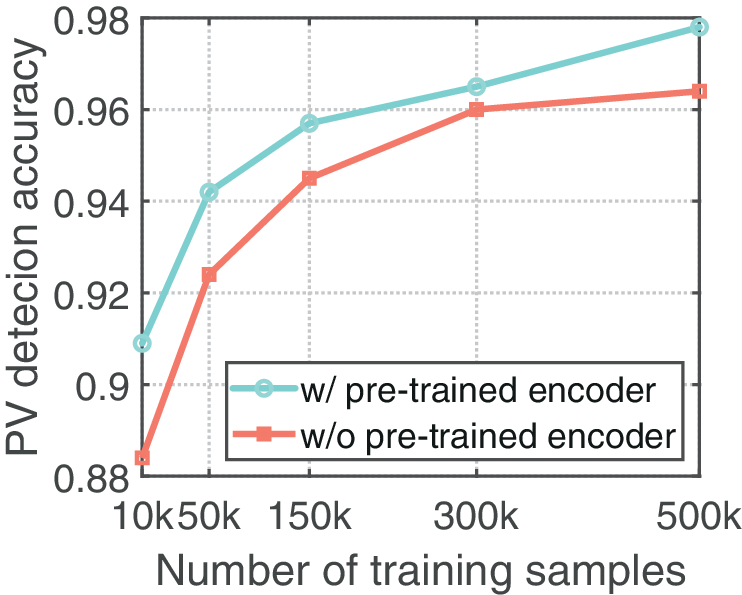}\label{pv_detection}}
\hspace{1 em}
\subfloat[]{\includegraphics[width=1.5in]{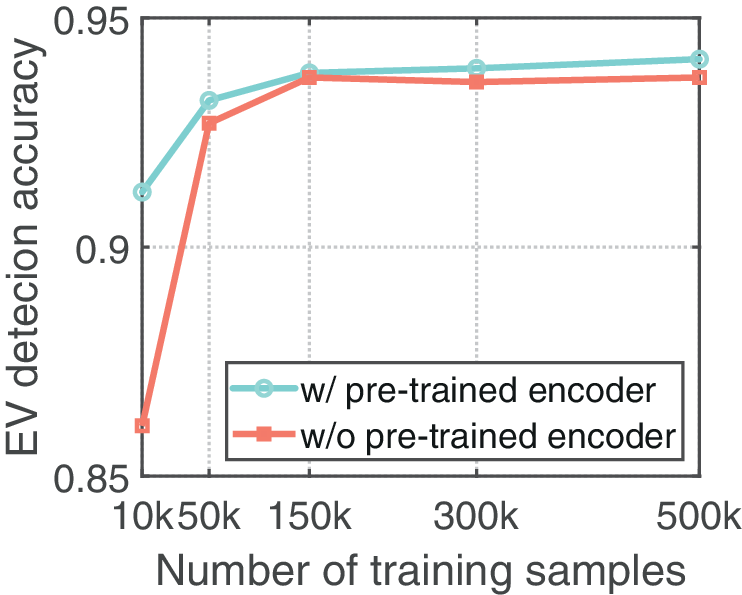}\label{ev_detection}}\\
\caption{Comparison between identification models with or without pre-trained encoder network. (a) PV generation identification, and (b) EV load identification.}
\label{effectiveness_pretrain}
%\vspace{-0.2in}
\end{figure}

As illustrated in Fig.~\ref{effectiveness_pretrain}, a significant performance boost is evident, particularly when the training dataset is limited, such as within the range of 10k to 150k examples, by incorporating the pre-trained ViT4LPA encoder. Even with the utilization of the complete dataset comprising 500k examples, a performance gain of 1-2\% is attainable when employing ViT4LPA. This outcome highlights that harnessing a pre-trained model not only effectively reduces the reliance on extensive training datasets but also consistently yields robust performance enhancements through comprehensive pre-training.

%The two sub-figures within Fig.~\ref{effectiveness_pretrain} illustrate the PV and EV detection accuracy for the test dataset, respectively, based on the number of available training samples. These results highlight the model's dependence on labeled data and the performance enhancement achieved through pre-training. Notably, when a limited amount of training examples is available (10k-150k), a substantial performance gap is observed between models with and without a pre-trained encoder. Even with the entire dataset (500k) accessible, a noticeable 1-2\% performance gain persists. This outcome affirms the pre-training model's ability to mitigate dataset dependency and achieve performance improvements for load image datasets.

%The third figure in Fig.~\ref{effectiveness_pretrain} compares overall detection accuracy on test data set by each training epoch. While the model without a pre-trained model converges to optimal network parameters after 50 epochs, the model with a pre-trained model converges around 25 epochs, supporting the effectiveness of training efficiency when leveraging a pre-trained model. 

In Table~\ref{Detection_benchmarkEVPV}, we compare the proposed ViT-based identification model with conventional CNN models. The CNN architectures used in our experiments are based on the Inception model~\cite{szegedy2015going}, with variations incorporating both 2D and 1D convolution layers, respectively.
The comparison reveals a significant performance gap between the proposed ViT-based model and the benchmark CNN models. While the PV identification accuracy appears similar across the models, a notable difference emerges in EV identification accuracy. For the PV identification task, the abundance of available data makes pre-training less impactful. However, when it comes to EV identification, labeled data is scarce, underscoring the increased importance of leveraging the knowledge acquired through pre-training for effective identification. Consequently, in the EV identification task, using ViT4LPA results in a substantial performance enhancement compared to using the two CNN models.
%The ViT4LPA can gain understanding of load patterns through the encoder network plays a crucial role in enhancing accuracy in the challenging context of EV detection.

%In Fig.~\ref{attn_map}, the attention maps of the cross-attention layer in the decoder network of the detection model are represented, and corresponding load images and daily load profiles of one day with sub-metering information are shown. One key finding is that each query tends to attend to local information in layer 1 and more global information in layer 2, which exhibits information flow to that of self-attention layers of the encoder represented in Fig~\ref{Encoder_interpret}. In layer 1, both queries attend to identical two patches, which represent the lowest power consumption period in the load image. One possible interpretation is that the model identifies the load pattern of the customer by analyzing the base load consumption level in the early layer. In layer 2, the PV detection query attends to most of the patches when solar irradiance is available and the EV detection query attends to evening to morning patches where the typical EV charging happens for residential customers. One possible detection mechanism is that each query compares the base load consumption information extracted in layer 1 and the load consumption pattern during possible PV generation or EV charging period extracted in layer 2. 

\begin{table}[h]
	\begin{center}
		\caption{Performance Comparison for Conducting PV and EV Load Identification Tasks}
		\label{Detection_benchmarkEVPV}
		\begin{tabular}{cccc}
	
			\toprule % <-- Toprule here
Model & Accuracy (PV) & Accuracy (EV) & Overall\\
\midrule
ViT (proposed) & 0.984 & \textbf{0.943} & \textbf{0.964}\\
Inception (2D) & \textbf{0.988} & 0.912& 0.950\\
Inception (1D) & 0.987 & 0.911& 0.949\\
\midrule
\bottomrule
		\end{tabular}
	\end{center}
\vspace{-0.2in}
\end{table}

\subsection{Downstream Task 2: HVAC Load Disaggregation}
Next, we showcase the performance improvement when using the pre-trained ViT4LPA encoder on load disaggregation tasks, using HVAC load disaggregation as an example. 
In Table~\ref{Detection_benchmark}, we compare ViT4LPA with two benchmark models: the Inception model and the bidirectional Long Short-Term Memory (BiLSTM) model. Note that BiLSTM is a RNN model widely employed for sequential data analysis. To evaluate the models, we use the normalized mean absolute error ($n$MAE) to assess point-to-point error and energy error (EE) to measure cumulative estimation deviation. For more details on the performance metrics used for load disaggregation, please refer to our prior work~\cite{ye2023modified}.

\vspace{-0.1in}
\begin{table}[h]
	\begin{center}
		\caption{Performance Comparison (Load Disaggregation Task)}
		\label{Detection_benchmark}
		\begin{tabular}{cccc}
	
			\toprule % <-- Toprule here
Model & $n$MAE (\%) & EE (kWh) & std($n$MAE)\\
\midrule
ViT (proposed) & \textbf{6.89} & \textbf{2.80} & \textbf{2.21}\\
Inception & 8.89 & 4.11& 2.59\\
BiLSTM & 8.01 & 3.53& 2.26\\
%Trans. decoder& 7.79 & 3.29& 2.23\\
\midrule
\bottomrule
		\end{tabular}
	\end{center}
\vspace{-0.2in}
\end{table}

\begin{figure}[h]
	\centering
	\includegraphics[width=2.8in]{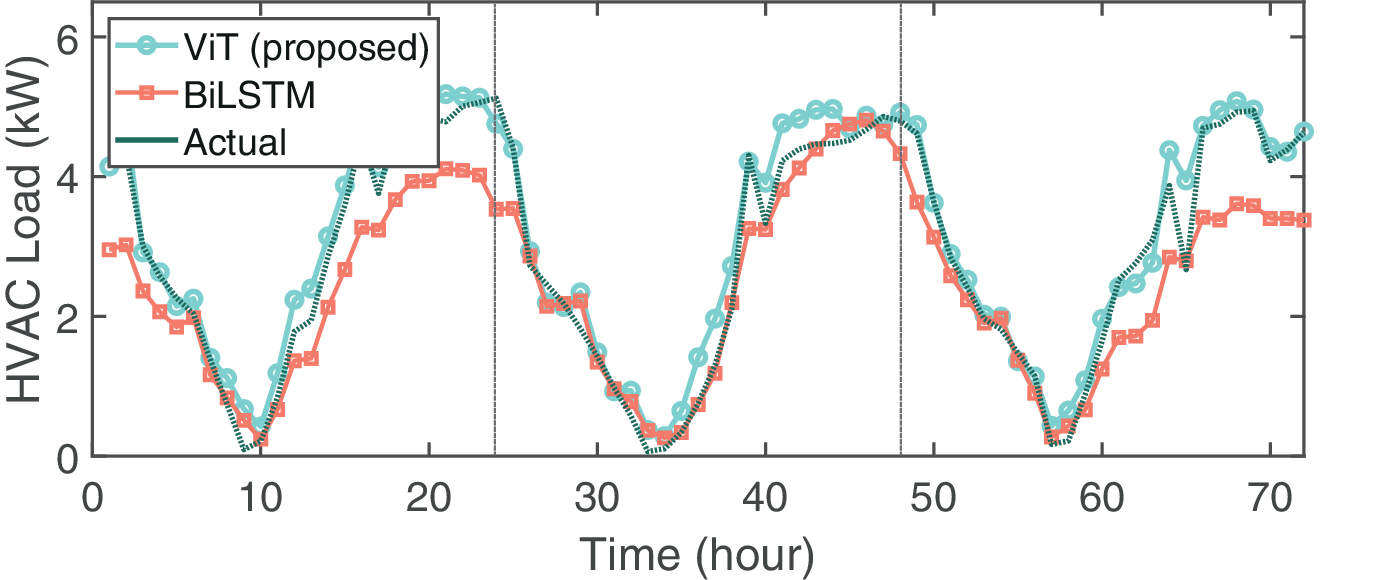}
\caption{Illustration of HVAC load disaggregation results over a three-day period for a customer.}
\label{disaggregation}
\vspace{-0.2in}
\end{figure}

\begin{figure}[h]
\centering
\subfloat[]{\includegraphics[width=1.3in]{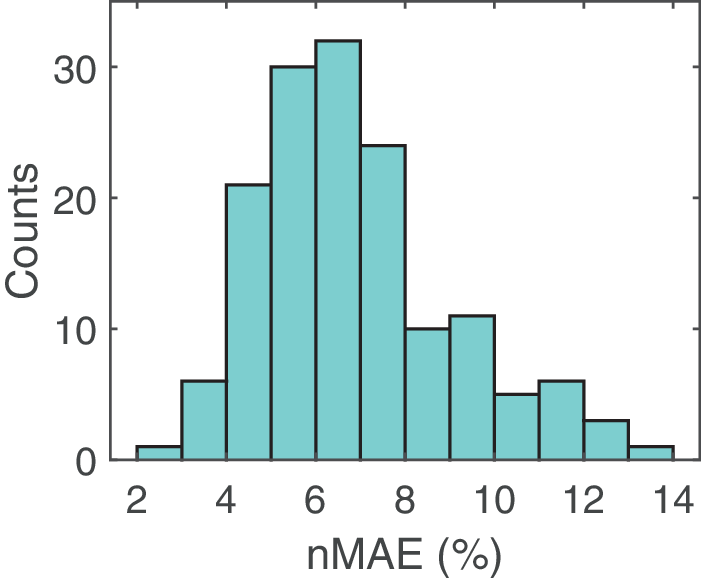}\label{nmae_vit}}
\hspace{1 em}
\subfloat[]{\includegraphics[width=1.3in]{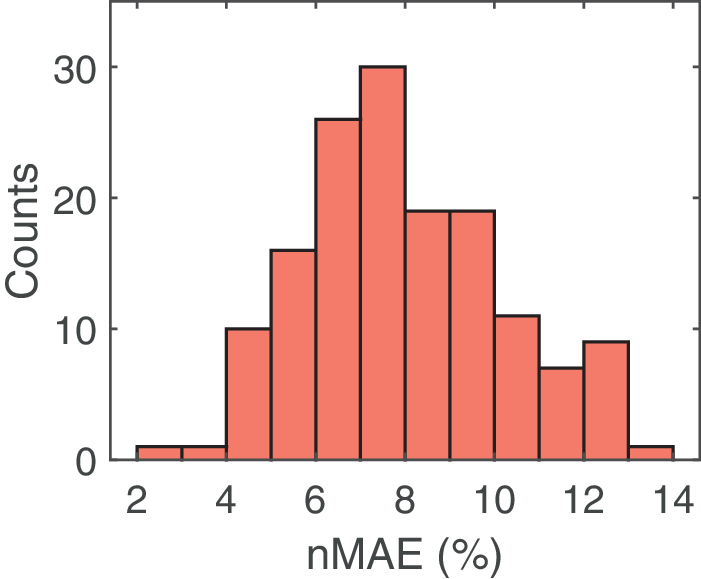}\label{nmae_lstm}}\\
\caption{Empirical distribution of $n$MAE disaggregation methods. (a) ViT4LPA model, and (b) BiLSTM model.
}
\label{distribution}
\vspace{-0.1in}
\end{figure}

From the results, we made the following observations
\begin{itemize}
    \item ViT4LPA exhibits the lowest point-to-point error, energy error, and error standard deviation.
    \item As shown in Fig.~\ref{disaggregation}, the HVAC load curve generated by ViT4LPA closely aligns with the ground truth, while the estimates derived from the BiLSTM model display noticeable errors, especially during peak HVAC load periods.
    \item As shown in Fig.~\ref{distribution}, ViT4LPA exhibits remarkable consistency, as evidenced by the narrow standard deviation of $n$MAE. It's important to note that each error observation is derived from averaging the resultant errors across multiple days for each customer, demonstrating the model's robustness across various customers. In the case of our proposed method, errors are densely concentrated within the 4-8\% range, indicating a high level of precision. Conversely, the errors of the BiLSTM models exhibit a comparatively broader distribution with a median value of 7.6\%.
    
\end{itemize}

%In Table~\ref{Detection_benchmark},  This superior performance is visually evident in Fig.~\ref{disaggregation}, where the HVAC profiles estimated by both models are compared with the ground truth measurements. The profiles generated by our proposed method closely align with the ground truth, showcasing minimal deviation. In contrast, the estimates derived from the BiLSTM model exhibit notable errors, particularly during peak HVAC load periods. The strength of our proposed model lies in its ability to harness customer-specific information extracted by the pre-trained encoder network. This advantage enables our model to accurately capture peak HVAC loads by analyzing load profiles spanning multiple days.

%Furthermore, our proposed model exhibits remarkable consistency, as evidenced by the narrow standard deviation of $n$MAE. This trend is also apparent in Fig.~\ref{distribution}, where the $n$MAE distribution of the two different models is depicted. It's important to note that each error observation is derived from averaging the resultant errors across multiple days for each customer, demonstrating the model's robustness across various customers. In the case of our proposed method, errors are densely concentrated within the 4-8\% range, indicating a high level of precision. Conversely, the errors of the BiLSTM models exhibit a comparatively broader distribution with a median value of 7.6\%.

\section{Conclusion}
In this study, we introduce ViT4LPA, a pre-trained Vision Transformer (ViT)-based load image encoder designed to generate load embeddings from load images. By converting load, temperature, and solar irradiance profiles into load images, we enable the direct adoption of ViT, a powerful image processing model, for Load Profile Analysis (LPA). Pre-trained on masked image restoration tasks, ViT4LPA captures correlations among load profile pixels to enhance performance in downstream tasks.
To gauge the performance of ViT4LPA, we conducted tests on two popular downstream LPA tasks: load identification and load disaggregation.
Our results demonstrate significant improvements in both tasks, with lower point-to-point errors, cumulative energy errors, and reduced standard deviations of errors. This underscores the potential of pre-trained models, particularly in machine learning and data analysis for power system applications. The capability of ViT4LPA to fine-tune with limited labeled data is especially valuable in applications constrained by data scarcity or concerns related to data privacy.

Moving forward, our research will delve into analyzing
suitable masking strategies and patch sizes to further enhance
the performance across multiple downstream tasks.

\bibliographystyle{IEEEtran}
\bibliography{conf}

\end{document}